\documentclass[conference]{IEEEtran}
\IEEEoverridecommandlockouts
\usepackage{cite}
\usepackage{amsmath,amssymb,amsfonts}
\usepackage{algorithmic}
\usepackage{graphicx}
\usepackage{textcomp}
\usepackage{xcolor}
\usepackage{url}
\usepackage{bm}
\usepackage{multirow}
\usepackage[inline]{enumitem}
    
\usepackage{glossaries-extra}

\setabbreviationstyle[acronym]{long-short}
\setabbreviationstyle{short-long}
\glssetnoexpandfield{long}
\glssetnoexpandfield{longpl}
\newcommand*{\parentabbr}[2]{%
  \ifglsused{#1}{#2\glsentryshort{#1}}{\protect\glsunset{#1}#2\glsentrylong{#1}}%
}

\newacronym{mtc}{MTC}{Machine Type Communications}
\newacronym{mmtc}{mMTC}{\parentabbr{mtc}{massive }}
\newacronym{m2m}{M2M}{Machine to Machine}
\newacronym{iot}{IoT}{Internet of Things}
\newacronym{gnss}{GNSS}{Global Navigation Satellite Systems}
\newacronym{esa}{ESA}{European Space Agency}
\newacronym{mcs}{MCS}{Modulation and Coding Scheme}
\newacronym{ra}{RA}{Random Access}
\newacronym{rv}{RV}{Random Variable}
\newacronym{pdf}{pdf}{probability density function}
\newacronym{gw}{GW}{Gateway}
\newacronym{meo}{MEO}{Medium Earth Orbit}
\newacronym{geo}{GEO}{Geostationary Earth Orbit}
\newacronym{leo}{LEO}{Low Earth Orbit}
\newacronym{vleo}{VLEO}{\parentabbr{leo}{very }}
\newacronym{gso}{GSO}{Geosynchronous Orbit}
\newacronym{ngso}{NGSO}{\parentabbr{gso}{Non }}
\newacronym{fov}{FOV}{Field Of View}
\newacronym{fsl}{FSL}{Free Space Loss}
\newacronym{ul}{UL}{Uplink}
\newacronym{dl}{DL}{Downlink}
\newacronym{snr}{SNR}{Signal-to-Noise Ratio}
\newacronym{sir}{SIR}{Signal-to-Interference Ratio}
\newacronym{sinr}{SINR}{Signal-to-Interference plus Noise}
\newacronym{awgn}{AWGN}{Additive White Gaussian Noise}
\newacronym{hpbw}{HPBW}{Half Power Beam Width}
\newacronym{nbiot}{NB-IoT}{NarrowBand-IoT}
\newacronym{b5g}{B5G}{Beyond 5G}
\newacronym{mmse}{MMSE}{Minimum Mean Square Error}
\newacronym{satcom}{SatCom}{Satellite Communications}
\newacronym{pac}{PAC}{Per Antenna Constraint}
\newacronym{mpc}{MPC}{Maximum Power Constraint}
\newacronym{csi}{CSI}{Channel State Information}
\newacronym{ssp}{SSP}{Sub Satellite Point}
\newacronym{cdf}{CDF}{Cumulative Distribution Function}

\def\BibTeX{{\rm B\kern-.05em{\sc i\kern-.025em b}\kern-.08em
    T\kern-.1667em\lower.7ex\hbox{E}\kern-.125emX}}
\begin{document}

\title{Beamforming in LEO Constellations for NB-IoT Services in 6G Communications}

\author{\IEEEauthorblockN{Alessandro Guidotti, Matteo Conti, Alessandro Vanelli-Coralli}
\IEEEauthorblockA{\textit{Department of Electrical, Electronic, and Information Engineering "Guglielmo Marconi" (DEI)} \\
\textit{University of Bologna}, Bologna, Italy \\
\{a.guidotti, matteo.conti19, alessandro.vanelli\}@unibo.it}
}

\maketitle

\begin{abstract}
With the first commercializations of 5G networks, \gls{b5g}, or 6G, systems are starting to be defined. In this context, \gls{iot} services will be even more impactful with respect to 5G systems. In order to cope with the huge amount of \gls{iot} devices, and the potentially large capacity requirements for the most advanced of them (\emph{e.g.}, live camera feeds from first responders in emergency scenarios), non-terrestrial systems will be pivotal to assist and complement the terrestrial networks. In this paper, we propose a multi-layer non-terrestrial architecture for \gls{nbiot} services based on \gls{gso} and \gls{ngso} nodes. To cope with both the number of devices and the potentially large capacities, we propose to implement \gls{mmse} beamforming in an aggressive full frequency reuse scenario. The performance assessment shows the significant benefits of the proposed solution.
\end{abstract}


\glsresetall

\section{Introduction}
While the vast potential of 5G networks is starting to drive the global economy and society, the race towards \gls{b5g}, or 6G, mobile communications has already started, \cite{ITU_FGNET_1,6G_flagship}. The evolution of 5G into 6G networks is aimed at responding to the increasing need of our society for ubiquitous and continuous connectivity services in all areas of our life: from education to finance, from politics to health, from entertainment to environment protection. The requested applications and services pose a vast variety of requirements calling for a flexible, adaptable, resilient, and cost efficient network able to serve heterogeneous devices with different capabilities and constraints. In this context, the technologies for 6G networks will include both novel concepts, as Terahertz communications and intelligent communication environments (\emph{e.g.}, intelligent surfaces), and the full exploitation of solutions initially adopted for 5G, \emph{e.g.}, Artificial Intelligence (AI) and \gls{satcom}, \cite{6G_paper_1,6G_paper_2,6G_paper_3}. 

\gls{iot}, as a widespread system interconnecting not only things, but also people, processes, and data, is already taking a central role in 6G. Recognising this, since Rel. 13, 3GPP introduced a number of key features to provide a progressively improved support to Low Power Wide Area Network (LPWAN) with EC-GSM-\gls{iot} and LTE-eMTC, \cite{3GPP_IOT_1,3GPP_IOT_2}. More recently, \gls{nbiot} has been introduced, aiming at, among the others, \cite{3GPP_NBIOT_1,3GPP_NBIOT_2}: 
\begin{enumerate*}[label=\roman*)]
	\item supporting ultra-low complexity devices for \gls{iot};
	\item improving the indoor coverage;
	\item supporting a massive number of connections ($52547$ per cell-site sector); and 
	\item improving the battery efficiency.
\end{enumerate*}
It shall be noticed that this is actually a new technology and, thus, not fully backward compatible with legacy 3GPP devices.
Based on the lessons learnt from the initial 5G deployments and on the global societal and network trends, the translation from human-centric towards machine-centric communications will be even more pronounced in 6G. In fact, it is expected that the number of Machine-To-Machine (M2M) connections will increase from 8.9 billions in 2020 to 14.7 billions in 2023, representing one third of the global devices, \cite{Cisco}. In addition, also the total traffic is going to intensify, motivated by the deployment of advanced services, \emph{e.g.}, telemedicine and smart transportation, which might require larger bandwidths compared to wearables and sensor networks. Both in terms of connections and bandwidth, terrestrial networks alone might not be able to satisfy the demanding \gls{iot} requirements in the near future. In addition, another vital requirement for an infrastructure providing \gls{iot} services is that of guaranteeing ubiquitous connectivity to the low-cost, low-powered devices distributed all over the globe. This again might be a challenge for terrestrial \gls{iot}, since there are vast areas where a terrestrial infrastructure is unfeasible or not economically viable.

In this framework, the integration of satellite and terrestrial networks can be a cornerstone to the realisation of the foreseen heterogeneous global 6G system. Thanks to their inherently large footprint, satellites can efficiently complement and extend dense terrestrial networks, both in densely populated areas and in rural zones, as well as provide reliable Mission Critical services; these aspects, and the related technical challenges, are being extensively addressed in academic research and funded projects, \cite{5G_paper_1,5G_paper_2,5G_paper_3,5G_paper_4,5G_paper_5,5G_paper_6,5G_paper_7}. The added value of a Non-Terrestrial Network (NTN) component in the overall system architecture has also been recognised by 3GPP, which officially approved two Work Items aimed at integrating NTN in the 5G architecture starting from Rel. 17, \cite{TR38821,TR38811}. In this context, several studies and funded projects are addressing the integration of satellite and terrestrial communications for \gls{iot} services, \cite{project_1,project_2,project_3}. In \cite{paper_1}, an overview of the \gls{iot} scenarios and use cases in which satellite systems can play a key role is provided. The authors of \cite{paper_2} focus on \gls{leo}-based \gls{iot} services, analyzing the possible use cases and proposing a  constellation design. In \cite{paper_3}, the authors study the LoRa adaptability in the \gls{iot} use case for \gls{leo} satellites. In \cite{paper_4}, the authors explore the potential of \gls{satcom} systems in 5G Machine Type Communications (MTC). In \cite{paper_5}, the authors provide a detailed analysis on the link budget calibration for \gls{nbiot} based on 3GPP specifications. An overview of the possible challenges of \gls{nbiot} over NTN is reported in \cite{paper_6}. An analysis on the Doppler and delay induced by non-geostationary satellites is reported in \cite{paper_7,paper_8,paper_9}. Finally, \cite{paper_10} addresses the design of a \gls{nbiot} receiver in the presence of Doppler effects at the gateway side.

In this paper, we focus on the implementation of beamforming techniques in \gls{ngso} constellations for \gls{nbiot} services. As previously introduced, the main challenges are related to providing ubiquitous connectivity also in remote areas and supporting a massive number of \gls{iot} devices, potentially with large bandwidth requirements. For these reasons non-terrestrial (NT) systems can provide a valuable resource for 6G \gls{iot} services to support and complement terrestrial networks. In this context, traditional frequency reuse approaches, \emph{e.g.}, 3 or 4 colours, might not be sufficient, thus calling for more aggressive solutions as full frequency reuse. To deal with the significantly increased interference due to the antenna side-lobes, advanced interference management techniques, as beamforming, precoding, and Multi-User Detection (MUD), are needed. During the last years, the implementation of beamforming techniques in \gls{satcom} has been extensively addressed for \gls{geo} systems, mainly, but also for \gls{leo} constellations, \cite{precoding_1,precoding_2,precoding_3,precoding_4,precoding_5,precoding_6} and references therein. In these works, the objective has been that of increasing the overall throughput in unicast or multicast systems, also addressing well-known issues for \gls{satcom}-based beamforming as scheduling and \gls{csi} retrieval. In this paper, we propose an advanced system architecture for NT-based \gls{iot} services and implement beamforming solutions to show the benefit that can be obtained for \gls{iot} services.

\begin{figure}[t]
	\centerline{\includegraphics[width=0.46\textwidth]{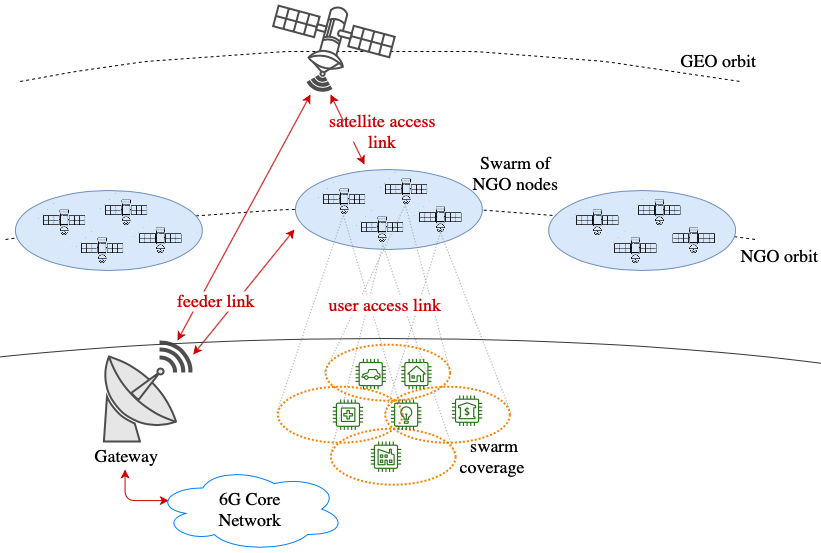}}
	\caption{Multi-layered non-terrestrial architecture for beamforming-based \gls{iot}.}
	\label{fig:architecture}
\end{figure}

\section{Non-terrestrial NB-IoT architecture}
Building on the multi-layered NT network introduced in \cite{5G_paper_3}, we propose the system architecture showed in Fig.~\ref{fig:architecture}:
\begin{itemize}
	\item The terrestrial segment is composed by the 6G terrestrial network infrastructure, which is connected to the NT segment through a set of on-ground \glspl{gw}. The \glspl{gw} have to be properly distributed around the globe so as to guarantee that at least one \gls{gw} is always in \gls{fov} of one of the elements in the non-terrestrial segment, \emph{i.e.}, a \gls{gso} or \gls{ngso} element. Each \gls{gw} manages a set of 6G base stations, which we refer to as \emph{future generation NodeB} (fgNB), to avoid confusion with 5G legacy gNBs; the fgNBs are thus conceptually located at the \gls{gw}. In this context, it shall be highlighted that: 
	\begin{enumerate*}[label=\roman*)]
	    \item with transparent payloads in the NT elements, the on-ground fgNB implements the full protocol stack; 
	    \item with regenerative payloads, functional split solutions can be considered, thus separating the fgNB protocol stack between the on-ground central unit (CU-fgNB) and the flying distributed unit (DU-fgNB). 
	\end{enumerate*}
	The implementation of functional split and the type of split depend on the complexity and capabilities of the NT segment.
	\item The access segment is assumed to be deployed as a multi-layer NT network, \emph{i.e.}, with elements located on different orbits, including \gls{gso} and \gls{ngso}. Focusing on the latter, depending on the considered altitude, various elements can be launched: \gls{leo} or \gls{vleo} satellites, CubeSats, High Altitude Platform Stations (HAPS), balloons, etc.; given the very diverse characteristics and performance they have, we generically refer to these elements as \emph{\gls{ngso} nodes}. As highlighted in Fig.~\ref{fig:architecture}, these nodes are connected to the terrestrial segment through a \gls{gso} satellite or by directly connecting to the visible \glspl{gw}. Each node is able to serve a portion of the on-ground coverage area by means of a single-beam or multi-beam system, depending on the payload. It shall be noticed that, for the sake of clarity, we represented one \gls{ngso} layer only, but more can be envisaged, based on a proper trade-off between the system requirements and its complexity. 
	\item In the on-ground user segment, a plethora of \gls{iot} devices is distributed all over the world and they are served by the \gls{ngso} nodes through the user access link. As previously mentioned, these devices potentially have heterogeneous requirements depending on the type of service. Notably, both one-way uplink and two-way \gls{iot} is possible. In the former case, the typical scenario is that of an enormous amount of \gls{iot} devices requiring both limited capacities, as location (\emph{e.g.}, smart transportation, automotive, good and freight tracking, etc.) and monitoring measurements (\emph{e.g.}, energy production in smart grids, farmland sprinkler irrigation), and larger ones, as live feeds from security cameras or first responders in an emergency scenario. In the latter case, there is a growing pervasiveness for bidirectional communications, including, but not limited to, wearable devices (\emph{e.g.}, answering to text or calls from a smart-watch), smart homes (\emph{e.g.}, remote control of connected devices), and Mission Critical scenarios (\emph{e.g.}, sending live feeds from the first responders to the control center). In addition, it shall also be considered that in many scenarios the distributed measurements from \gls{iot} sensors are collected in a centralised unit that controls and optimises the system behaviour by reconfiguring the actuators (which can be embedded in the same sensing \gls{iot} device or separately), \emph{e.g.}, optimisation at city or even regional level of a smart energy distribution grid, \cite{smart_grid_1,smart_grid_2}.
\end{itemize}

The type of connection between the terrestrial and the access segments depends on both the complexity of the on-ground network, \emph{i.e.}, the number of \glspl{gw} that can be deployed, and the position of the \gls{ngso} nodes on the orbit; in fact, it will happen during one orbit period that the nodes are moving over areas where the visibility with a \gls{gw} might be difficult (very low elevation angles) or not possible at all and, thus, the connection with a \gls{gso} satellite shall be guaranteed to ensure communications with the core network. Clearly, where possible, both connections (\gls{gw} and \gls{gso}) can be considered to cooperatively work. With respect to the air interface to be used on these links, the type of payload, \emph{i.e.}, transparent or regenerative, and the possible presence of functional split have to be considered. To better detail this aspect, it is reasonable to assume that similar approaches in the standardisation will be used for 6G as those implemented in 5G. Thus: 
\begin{enumerate*}[label=\roman*)]
	\item with a transparent payload, the connection between the \glspl{gw} and the \gls{ngso} nodes, as well as that between \gls{gso} and \gls{ngso}, shall be based on the same air interface of the user access link, discussed below; 
	\item with regenerative payloads, the air interface shall be that between a fgNB and the core network without functional split (NG air interface in 5G systems) or that interconnecting the central and distributed units of the fgNB (F1 interface in 5G). 
\end{enumerate*}
It is worth mentioning that, in the latter case, the NG and F1 interfaces can be assumed to be logical ones as for legacy 5G systems, \emph{i.e.}, as long as proper signalling is ensured, they can be implemented by means of any Satellite Radio Interface (SRI), \emph{e.g.}, DVB-S2, \cite{S2}, DVB-S2X, \cite{S2X}, or DVB-RCS, \cite{RCS}.

In terms of the user access link, it shall be highlighted that the on-ground \gls{iot} devices do not have visibility of all of the \gls{ngso} nodes for the entire time. In particular, depending on the nodes' location along the orbit, for a given on-ground coverage area, only a subset of nodes will be able to communicate with the \gls{iot} devices. Based on this observation, the \gls{ngso} nodes are grouped into \emph{\gls{ngso} swarms}, \emph{i.e.}, groups of \gls{ngso} nodes that are serving, through their combined single-/multi-beam footprints, a specific on-ground area for the current orbital position. The previously discussed connection between the \gls{ngso} nodes and the core network is to be thus intended as a connection at swarm level; in this context, in case a single node of the swarm is able to connect to a \gls{gw}, either the whole swarm is also served through the proposed \gls{gso} satellite or Inter-Node Links (INLs) have to be considered to ensure proper coordination.  If not otherwise specified, we assume the 3GPP \gls{nbiot} air interface on the user access link without affecting the generality of our work; however, other options are also possible, \emph{e.g.}, SigFox or LoRa.

\section{System Model}
In the proposed multi-layer non-terrestrial system, we focus on the user access downlink towards the \gls{nbiot} devices. If not otherwise specified, in the following we assume that: 
\begin{enumerate*}[label=\roman*)]
	\item a swarm of $N_{s}$ single-beam \gls{ngso} nodes is serving the on-ground \gls{iot} devices based on a full frequency reuse scheme; 
	\item on-ground \gls{mmse} beamforming is implemented at the system \gls{gw}, which is assumed to ideally know the \gls{nbiot} \gls{csi}; and
	\item in the time-frequency resource grid, which is licit to be assumed for 6G, during each time slot, $N_s$ devices are randomly scheduled in each $180$ kHz \gls{nbiot} carrier, \emph{i.e.}, one device per beam. 
\end{enumerate*}
Focusing on a generic time slot, the signal received by the $i$-th device of the $b$-th beam can be written as:
\begin{equation}
\label{eq:rx_signal}
y_{b}^{(i)} = \sqrt{P_t}\mathbf{h}_{b}^{(i)}\mathbf{W}_{:,b}x_b + \sqrt{P_t}\sum_{\substack{\ell=1\\ \ell\neq b}}^{N_s}\mathbf{h}_{b}^{(i)}\mathbf{W}_{:,\ell}x_{\ell} + z_b^{(i)}
\end{equation}
where: 
\begin{enumerate*}[label=\roman*)]
	\item $\mathbf{x}=\left\{x_j\right\}_{j=1,\ldots,N_s}$, is the vector of complex unit-variance transmitted symbols; 
	\item $P_t$ is the maximum transmission power from each \gls{ngso} node;
	\item $z_b^{(i)}$ is a complex circularly-symmetric Gaussian random variable with zero-mean and unit variance; and 
	\item $\mathbf{h}_{b}^{(i)}$ represents the channel vector of the $i$-th device in the $b$-th beam. 
\end{enumerate*}
The channel coefficient between the $i$-th user in the $b$-th beam and the $j$-th satellite is:
\begin{equation}
	h_{b,j}^{(i)} = \frac{\sqrt{G_{T_{b,j}}^{(i)}G_{R_{b,j}}^{(i)}}}{4\pi\frac{d_{b,j}^{(i)}}{\lambda}\sqrt{A_{loss}P_z}}e^{-\jmath\frac{2\pi}{\lambda}d_{b,j}^{(i)}},\ j=1,\ldots,N_s
\end{equation}
in which: 
\begin{enumerate*}[label=\roman*)]
	\item $G_{T_{b,j}}^{(i)}$ and $G_{R_{b,j}}^{(i)}$ represent the transmitting and receiving antenna gains between the $j$-th satellite and the $i$-th user in the $b$-th beam;
	\item the transmitting antenna gain is a function of $\vartheta_{T_{b,j}}^{(i)}$, the angle between the $j$-th satellite antenna boresight and the direction of the $i$-th user in the $b$-th beam;
	\item the receiving antenna gain is a function of $\vartheta_{R_{b,j}}^{(i)}$, the angle between the antenna boresight of the $i$-th user in the $b$-th beam and the $j$-th satellite; 
	\item $A_{loss}$ models the antenna losses;
	\item $d_{b,j}^{(i)}$ is the slant range between the $i$-th user in the $b$-th beam and the $j$-th satellite;
	\item $\lambda$ is the carrier wavelength; and
	\item $P_z$ is the noise power. 
\end{enumerate*}
Please note that the unit variance assumption for the Gaussian noise in (\ref{eq:rx_signal}) is licit since the channel coefficients are normalised to the noise power $P_z$. Finally, $\mathbf{W}$ is the $N_s\times N_s$ \gls{mmse} beamforming matrix computed as:
\begin{equation}
\label{eq:beamforming}
	\bm{\mathrm{W}} = {\left({\bm{\mathrm{H}}}^H\bm{\mathrm{H}} + \mathrm{diag}\left(\bm{\alpha}\right)\bm{\mathrm{I}}_{\mathit{N_B}}\right)}^{-1}{\bm{\mathrm{H}}}^H
\end{equation}
where $\mathrm{diag}\left(\bm{\alpha}\right)$ is the vector of regularisation factors, with $\alpha_b=1/P_{t}$ $\forall b=1,\ldots,N_s$, \cite{optimal_alpha}. The above \gls{mmse} beamforming matrix is computed at each time slot based on the devices to be served. It is worth highlighting that specific power constraints are usually taken into account when computing the precoding matrix; these are aimed at regularising the combination of the transmitted power level from each on-board antenna so as to satisfy power constraint requirements, \cite{precoding_6}. In the considered system, we are implementing \gls{mmse} by means of a distributed antenna system, in which the antennas are located on-board the \gls{ngso} nodes; thus, rather than typical constraints focusing on the limitation of the overall power, we consider those in which the power per transmitting satellite is upper-bounded to $P_t$:
\begin{itemize}
	\item \gls{pac}: the maximum power per satellite is fixed to $P_t$, but the orthogonality among the \gls{mmse} matrix columns is degraded, leading to a loss in the performance due to an increased interference level.
	\begin{equation}
	\label{eq:pac}	
	\widetilde{\bm{\mathrm{W}}} = \frac{1}{\sqrt{N_B}}\mathrm{diag}\left(\frac{1}{\parallel {\bm{\mathrm{W}}}_{{1},:} \parallel },\ldots,\frac{1}{\parallel \bm{\mathrm{W}}_{\mathit{N_B},:} \parallel }\right)\bm{\mathrm{W}}
	\end{equation}
	\item \gls{mpc}: the power per satellite is upper bounded to $P_t$, but the orthogonality among the \gls{mmse} matrix columns is kept; in this case, one satellite is exploiting the maximum power $P_t$, while the others are transmitting lower power levels.
	\begin{equation}
	\label{eq:mpc}
	\widetilde{\bm{\mathrm{W}}} = \frac{\bm{\mathrm{W}}}{\sqrt{{N_B\max_{j}{\parallel\bm{\mathrm{W}}_{\mathit{j},:} \parallel}^2}}}
	\end{equation}
\end{itemize}
Based on the above considerations and assumptions, the \gls{sinr} at the $i$-th device in the $b$-th beam can be computed as:
\begin{equation}
\label{eq:sinr}
\gamma_b^{(i)} = \frac{P_{TX}{\left|\bm{\mathrm{h}}_b^{(i)}{\widetilde{\bm{\mathrm{W}}}}_{:,b}\right|}^2}{1+P_{TX}\sum_{\ell\neq b}{\left|\bm{\mathrm{h}}_b^{(i)}{\widetilde{\bm{\mathrm{W}}}}_{:,\ell}\right|}^2}
\end{equation}
From this, the achieved rate can be evaluated either from the Shannon formula or from the adopted Modulation and Coding (ModCod) scheme. It is worth highlighting that the above model is valid independently from the type of \gls{ngso} node that is considered. Moreover, even though we are focusing on the \gls{mmse} approach, other beamforming algorithms or normalisations can be easily included.


\section{Numerical Results}
While the proposed architecture is valid for a node constellation with global coverage, in the following we focus on a single \gls{ngso} swarm of $7$ \gls{leo} satellites in circular polar orbit. Each satellite is equipped with a single-beam antenna pointed towards the corresponding \gls{ssp}; we are thus assuming the moving beams scenario, \cite{TR38821}, \emph{i.e.}, the beams move together with the satellites along their orbit. For the swarm design, the following procedure is implemented: 
\begin{enumerate*}[label=\arabic*)]
    \item $\vartheta_{3dB}$, \emph{i.e.}, the angle at which the transmission radiation pattern is equal to $-3$ dB, for the considered antenna configuration is computed; 
    \item the beam radius is obtained in $uv$-coordinates as $\sin\vartheta_{3dB}$;
    \item an hexagonal lattice is generated for the \glspl{ssp}; 
    \item based on Earth-satellite geometry considerations, the inter- and intra-node separation angles, among orbits and satellites in the same orbit, respectively, are obtained for the considered altitude, providing the needed orbital parameters for the swarm design.
\end{enumerate*}

\begin{table}[t!]
	\caption{Simulation parameters: \gls{nbiot} device.}
	\begin{center}
		\begin{tabular}{|c|c|c|}
			\hline
			\textbf{Parameter} & \textbf{Value} & \textbf{Units} \\
			\hline
			antenna model & NTN, \cite{TR38811} & - \\
			\hline
			$G_R$ & 0 & dBi \\
			\hline
			antenna temperature & $290$ & K\\
			\hline
			noise figure & $7$ & dB\\
			\hline
		\end{tabular}
		\label{tab:IOT_parameters}
	\end{center}
\end{table}

\begin{table}[t!]
	\caption{Simulation parameters: \gls{leo} satellites.}
	\begin{center}
		\begin{tabular}{|c|c|c|c|c|c|c|}
			\hline
			\multirow{2}{*}{\textbf{Parameter}} & \multicolumn{2}{c|}{$600$ km} & \multicolumn{2}{c|}{$1200$ km} & \multirow{2}{*}{\textbf{Units}} \\
			\cline{2-5}
			 & \textbf{Set a} & \textbf{Set b} & \textbf{Set a} & \textbf{Set b}  & \\
			\hline
			antenna diameter & $2$ & $1$ & $2$ & $1$ & m\\
			\hline
			$G_{T,max}$ & $30$ & $24.1$ & $30$ & $24.1$ & dBi\\
			\hline
			EIRP density & $34$ & $30.3$ & $40$ & $35.7$ & dBW/MHz\\
			\hline
			$N_s$ & \multicolumn{4}{c|}{$7$} & -\\
			\hline
		\end{tabular}
		\label{tab:sat_parameters}
	\end{center}
\end{table}

The simulation parameters are reported in Tables \ref{tab:IOT_parameters} and \ref{tab:sat_parameters} for the \gls{nbiot} devices and the \gls{leo} satellites, respectively. We assume a system operating in S-band at $2$ GHz, in which, as per 3GPP specifications, $30$ MHz can be assumed as channel bandwidth; thus, exploiting a full frequency reuse approach with beamforming, approximately $166$ \gls{nbiot} devices can be simultaneously served with Frequency Division Multiplexing (FDM). As for the \gls{nbiot} devices, it is worth mentioning that omni-directional antennas are considered, \cite{IOT_parameters}, and, thus, $G_{R_{b,j}}^{(i)} = G_R$, $\forall i,b,j$. As for the losses modelled in $A_{loss}$, the following terms are included, \cite{TR38821}:
\begin{enumerate*}[label=\roman*)]
	\item the \gls{nbiot} devices are linearly polarised, \cite{IOT_parameters}, and, thus, an additional $3$ dB loss is considered; \item $2.2$ dB are introduced by ionospheric scintillations; and
	\item $3$ dB are included as shadowing margin.
\end{enumerate*}
As for the location of the \gls{nbiot} devices, at each of the $500$ Monte Carlo iterations, we assume them to be uniformly distributed with a density $\rho=1$ user/km$^2$; at each iteration, all of the considered users are served. In this context, it is worth to highlight that the user density is below the values expected in densely populated areas, while it is reasonable for remote regions; however, since we are assuming moving beams and considering ideal \gls{csi} at the \glspl{gw}, this assumption does not affect the generality of the numerical results: the transmission on each carrier is a snapshot at the current swarm location of a uniform distribution of \gls{nbiot} devices.
\begin{figure}[t]
	\centerline{\includegraphics[width=0.46\textwidth]{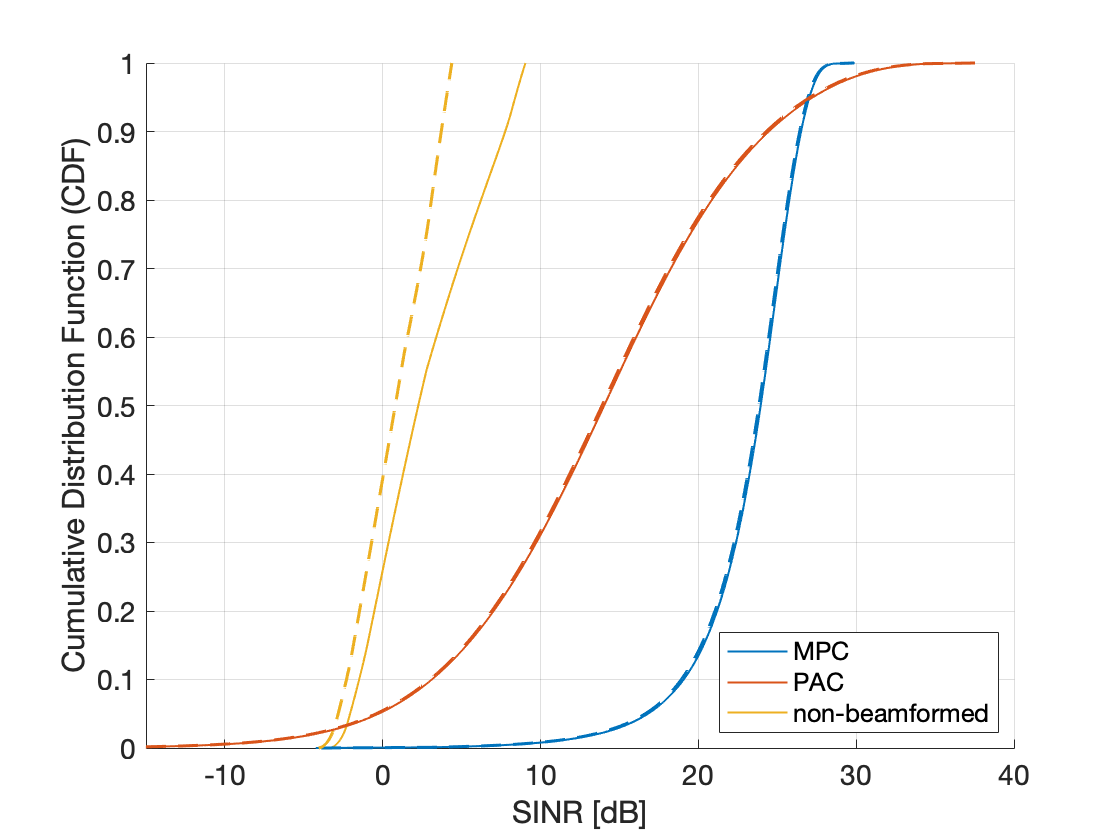}}
	\caption{SINR CDF for the set a configuration at $600$ km (solid) and $1200$ km (dashed).}
	\label{fig:set_a_cdf}
\end{figure}
\begin{figure}[t]
	\centerline{\includegraphics[width=0.46\textwidth]{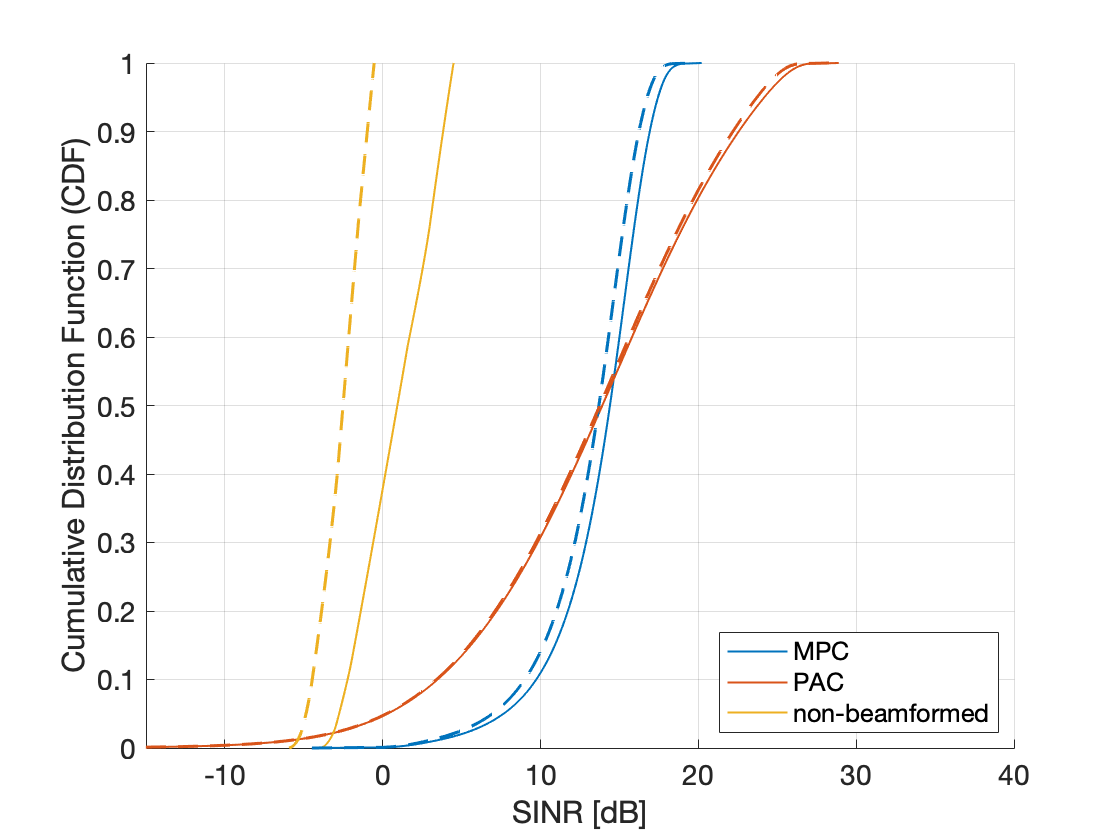}}
	\caption{SINR CDF for the set b configuration at $600$ km (solid) and $1200$ km (dashed).}
	\label{fig:set_b_cdf}
\end{figure}
\begin{figure}[t]
	\centerline{\includegraphics[width=0.46\textwidth]{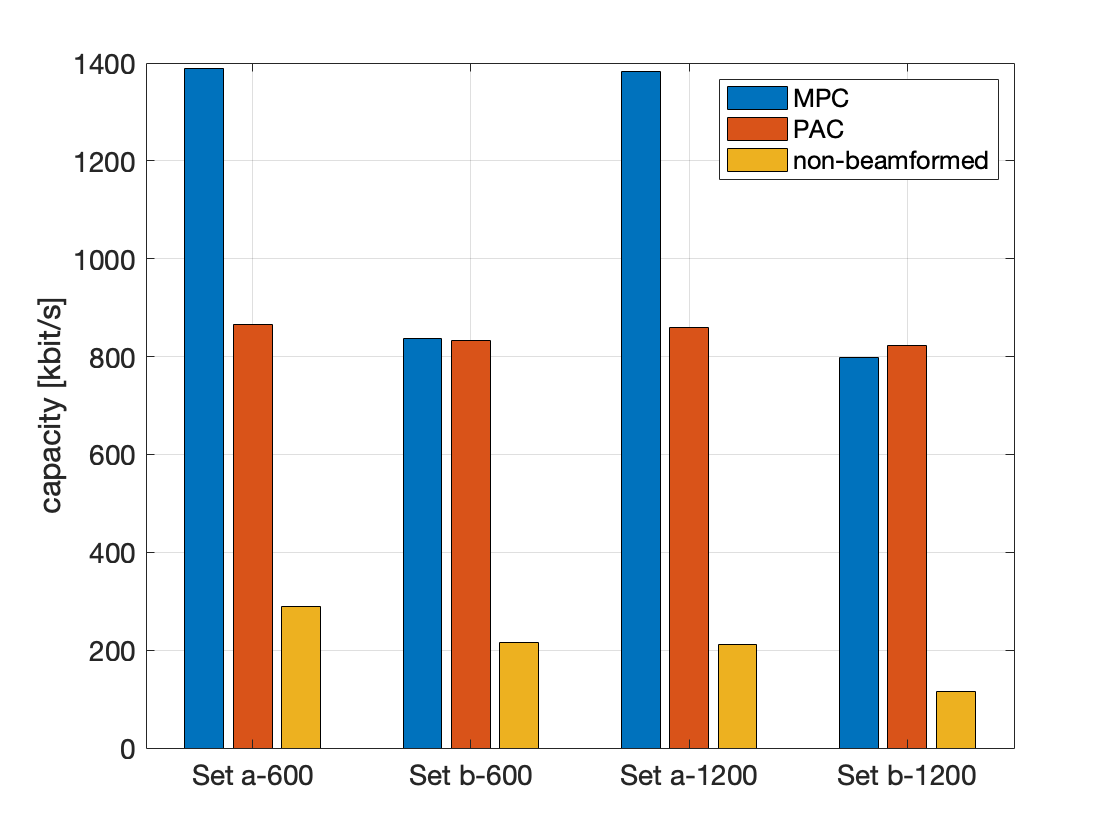}}
	\caption{Average capacity per NB-IoT device.}
	\label{fig:capacity}
\end{figure}
\subsection{Performance assessment}
In Fig.~\ref{fig:set_a_cdf} and \ref{fig:set_b_cdf} we report the \gls{cdf} of the Shannon spectral efficiency per user. Focusing on \textit{set a} (Fig.~\ref{fig:set_a_cdf}), we can notice that: 
\begin{enumerate*}[label=\roman*)]
	\item the performance with beamforming is significantly better compared to the non-beamformed case, with \gls{sinr} values that go up to $23-25$ dB for $90\%$ in the former case, compared to less than $10$ dB in the latter;
	\item the performance of \gls{mpc} is better in terms of fairness, with a \gls{cdf} slope much steeper than \gls{pac}; 
	\item with \gls{pac}, the maximum achievable \gls{sinr} is slightly better; and
	\item the performance with beamforming is almost identical for both $600$ and $1200$ km orbits.
\end{enumerate*} 
When \textit{set b} (Fig.~\ref{fig:set_b_cdf}) is considered, with less directive antennas, similar considerations hold; the only difference is that the performance at $1200$ km is slightly worse with respect to $600$ km orbits, in the order of less than $1$ dB. Comparing the performance with \textit{set a} and \textit{set b}, \gls{mpc} provides a better performance with \textit{set a}; this is motivated by observing that this set has a larger antenna gain; thus, since with \gls{mpc} onl one satellite is transmitting at $P_t$, the lower  transmission power of the other satellites that keeps the optimality of the \gls{mmse} solution is compensated.\\
In Fig.~\ref{fig:capacity}, the average capacity in the scenarios detailed in Table~\ref{tab:sat_parameters}, assuming one \gls{nbiot} carrier per user, \emph{i.e.}, $180$ kHz per user; it shall be noticed that, in case larger capacities are requested from some devices (as discussed in the previous sections), more carriers can be assigned by the system scheduler. It can be noticed that the trends highlighted in the \gls{sinr} \glspl{cdf} are confirmed: 
\begin{enumerate*}
	\item the \gls{mpc} normalisation provides a significantly better performance with respect to \gls{pac} at $600$ km ($1400$ vs $800$ kbit/s), while at $1200$ km they are similar;
	\item the benefit of beamforming is significant, with gains up to $380\%$ for \gls{mpc} at $600$ km. 
\end{enumerate*}
Comparing the performance with set a and set b, it can be noticed that \gls{mpc} is significantly better at $600$ km compared to \gls{pac}, while at $1200$ km they are almost identical.  Finally, it shall be noticed that these capacity values are sufficient for low-capacity \gls{nbiot} devices, but in case larger capacities are needed, more carriers can be aggregated.

\subsection{Technical challenges and future works}
The numerical assessment discussed above provides a valuable insight on the benefit that beamforming techniques can bring into future 6G \gls{nbiot} services. However, there are several technical challenges that still need to be addressed in future extensions of this work:
\begin{itemize}
	\item The numerical assessment has been performed with ideal \gls{csi} and considering a single swarm. More accurate analyses have to be performed assuming non-ideal \gls{csi}, \emph{i.e.}, by taking into account the capabilities of the \gls{nbiot} devices to estimate the channel vectors and to provide these measurements to the \glspl{gw}, which poses power consumption challenges due to both the estimation and the transmission of the \glspl{csi} back to the \glspl{gw}; the consideration of non-ideal \gls{csi} also allows to take into account the fact that, due to the swarm high speed on low orbits, those used at the \gls{gw} are likely to be outdated with respect to the actual channel condition. Moreover, multiple swarms will be considered to also take into account the increased interference coming from adjacent \gls{ngso} nodes in other swarms.
	\item The \gls{ngso} swarm constellation has been designed assuming the central node on the equatorial line; when the nodes move to different latitudes, due to the circular polar orbit characteristics, they will tend to be closer to each other and, thus, an increased overlap of the on-ground footprint will arise. This aspect has to be taken into account since it affects the beamforming performance.
	\item In this work, we focused on the downlink for two-way \gls{iot} services. However, also uplink transmissions pose challenges in terms of number of connections and potentially large bandwidth. In this context, uplink Cooperative MultiPoint (CoMP) techniques or distributed on-ground solutions as crowd-sourcing approaches.
	\item The user distribution in \gls{iot} services can be strongly non-uniform; in particular, in densely populated areas thousands of devices can be present per square km, while in rural areas there are likely to me much less devices.
\end{itemize} 
These aspects will be thoroughly addressed in future extensions of this work.

\section{Conclusions}
In this paper, we proposed a multi-layer non-terrestrial architecture for \gls{nbiot} services based on \gls{gso} and \gls{ngso} nodes. To deal with the huge number of NB-IoT devices and the potentially large bandwidth requirements, MMSE beamforming has been proposed. The performance assessment has been discussed in terms of both \gls{sinr} \gls{cdf} and average capacity, showing that \gls{mmse} beamforming with different normalisation can bring significant benefits. Moreover, the main technical challenges related to the proposed architecture, including non-ideal \gls{csi} and full global coverage, have been detailed and will be addressed in future works.

\end{document}